\documentclass{article}

\usepackage[preprint,nonatbib]{neurips_2023}

\usepackage[utf8]{inputenc} 
\usepackage[T1]{fontenc}    
\usepackage{hyperref}       
\usepackage{url}            
\usepackage{booktabs}       
\usepackage{amsfonts}       
\usepackage{nicefrac}       
\usepackage{microtype}      
\usepackage{xcolor}         
\usepackage{graphicx}  
\usepackage{setspace}
\usepackage{multirow}
\usepackage{hyperref}

\title{Neighbor-Environment Observer: An Intelligent Agent for Immersive Working Companionship}

\author{%
  Zhe Sun, Qixuan Liang, Meng Wang, Zhenliang Zhang\thanks{Corresponding author.} \\
National Key Laboratory of General Artificial Intelligence, BIGAI\\%
\texttt{\{sunzhe, liangqixuan, wangmeng, zlzhang\}@bigai.ai} \\
}

\begin{document}

\maketitle

\begin{abstract}
Human-computer symbiosis is a crucial direction for the development of artificial intelligence. As intelligent systems become increasingly prevalent in our work and personal lives, it is important to develop strategies to support users across physical and virtual environments. While technological advances in personal digital devices, such as personal computers and virtual reality devices, can provide immersive experiences, they can also disrupt users' awareness of their surroundings and enhance the frustration caused by disturbances. In this paper, we propose a joint observation strategy for artificial agents to support users across virtual and physical environments. We introduce a prototype system, neighbor-environment observer (NEO), that utilizes non-invasive sensors to assist users in dealing with disruptions to their immersive experience. System experiments evaluate NEO from different perspectives and demonstrate the effectiveness of the joint observation strategy. A user study is conducted to evaluate its usability. The results show that NEO could lessen users' workload with the learned user preference. We suggest that the proposed strategy can be applied to various smart home scenarios.
\end{abstract}

\section{Introduction}

Human-computer symbiosis is one of the potential outcomes of human-computer interaction (HCI). 
Licklider envisioned a partnership between humans and computers that could be viable, productive, and thriving~\cite{licklider1960man}. 
Numerous efforts have been made from different perspectives to move closer to human-computer symbiosis, including computer vision~\cite{jaderberg2015spatial, Li_2018_CVPR, tu2002image, vaswani2017attention, krizhevsky2017imagenet}, 
natural language processing~\cite{bengio2000neural, collobert2011natural, mikolov2013distributed, bahdanau2014neural}, 
robotics~\cite{thrun2002probabilistic, lavalle2006planning, hafner2019learning}, 
and interdisciplinary fields~\cite{shen2018transdisciplinary, shrestha2019review, devlin2018bert}.
With the development of artificial intelligence (AI), remarkable achievements have emerged, such as deep learning (DL)~\cite{lecun2015deep} and reinforcement learning (RL)~\cite{sutton2018reinforcement}. 
Intelligent agents now possess the capability to detect objects, recognize people's identities, comprehend sentences, and even offer human-like feedback through audio or visual avatars.

\begin{figure*}
  \centering
  \includegraphics[width=\linewidth]{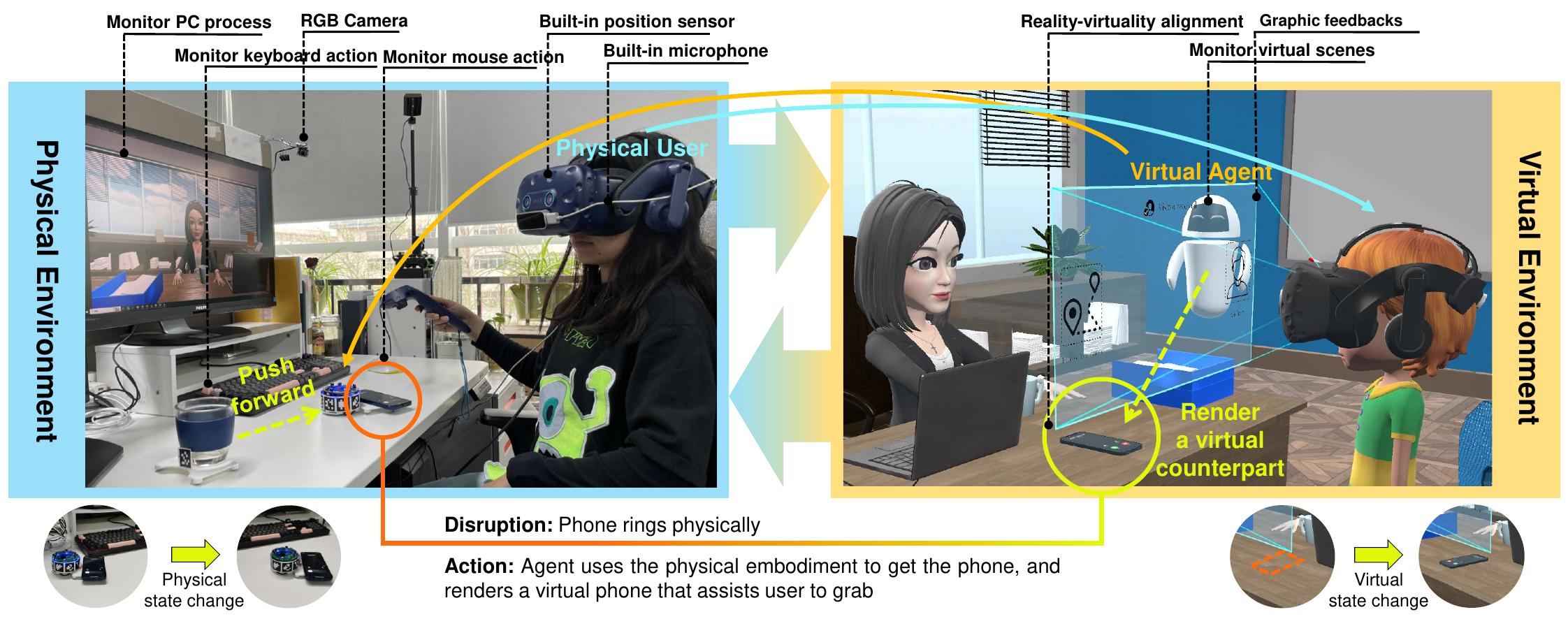}
  \caption{The NEO system jointly observes the physical and virtual environments and takes action in the two environments simultaneously. It is designed to deal with disruptions for people when they are immersed in a virtual working environment.} 
  \label{fig:teaser}
\end{figure*}

On the other hand, to approach an immersive working environment with digital content and agents, interface technologies such as virtual reality (VR), augmented reality (AR), and mixed reality (MR) have been developed. These technologies provide people with rich visual experiences and novel interactions~\cite{jones2020vroom}. When people are immersed in virtual environments, their virtual states may change vastly. So do their physical states. In such cases, cooperating with humans requires awareness not only of their states in the virtual environments but also in the physical environment. This can benefit from a combination of digital AI agents and smart-home agents.

There has been extensive research exploring digital AI agents for VR, AR, and MR environments. 
One notable approach is embodied artificial intelligence~\cite{gupta2021embodied}, which is designed to learn and explore in virtual simulated environments. 
With virtual embodiments, agents explore the virtual environment using not only visual information but also knowledge gained from interaction. 
This enables them to perform various tasks in VR, from point navigation~\cite{gordon2019splitnet} to simple daily interactions~\cite{srivastava2022behavior}. 
Consequently, agents based on embodied AI are more suitable for cooperating with humans by modeling human users' minds, which has further catalyzed research in the area of multi-agent systems~\cite{shum2019theory, jain2020cordial}.

However, human users do not solely stay in virtual environments, even when using VR Head-Mounted Devices (HMDs).
They often transition between the physical and virtual environments or receive information from both simultaneously.
For example, one still hears sounds and smells scents from the physical room around them when using VR HMDs.
This phenomenon of transition can be a source of disruption and can even lead to VR sickness~\cite{chang2020virtual}.
External stimuli, such as sudden temperature changes or gusts of wind, can disrupt immersion in the virtual world~\cite{tao2022integrating}, as can unexpected visitors at the door.
These changes in physical states are often not considered in traditional immersive systems but can be observed and intervened by an intelligent agent equipped with sensors and actuators. 
In this case, here comes the research question:

\begin{itemize}
    \item[RQ:] How can an artificial agent observe, understand, and cooperate with humans when they are immersed in the virtual environment, particularly in the presence of disruptions from the physical world?
\end{itemize}

Researchers have been examining the question of how do virtual agents effectively operate in both physical and virtual environments, and one perspective is the concept of symmetrical reality (SR), as proposed by Zhang et al.~\cite{zhang2018inverse, zhang2019symmetrical, zhang2021symmetrical, zhang2023building}. The SR paradigm involves enabling virtual agents to perceive and act in both environments.
Our work inherits the core concepts of SR paradigm and extends it in terms of technical implementation, prototype frameworks, and application cases.
An example is shown in \autoref{fig:teaser}.
In particular, we define the physical and virtual environments as ``neighbor environments'', as any digital device can serve as a gateway between them.
We formulated the following research questions to further delineate the overarching research question \textit{RQ}:
\begin{itemize}
    \item [RQ-1:] What methodology should be employed for an agent to understand users' states when they are in physical and virtual environments simultaneously?
    \item [RQ-2:] How can an agent align and parse information from both users and the environments in a coherent manner, and generate decisions autonomously without explicit commands from the users?
    \item [RQ-3:] What types of embodiments are necessary for the agent to execute decisions that involve both the physical and virtual environment?
    \item [RQ-4:] What design strategy of the agent's interface could reduce the users' disruption and enhance immersion?
\end{itemize}

These research questions guide our investigation into the underlying factors that are critical to the development of an effective and efficient agent that can seamlessly operate in both physical and virtual environments. 
By addressing these questions, we hope to advance the state-of-the-art in human-agent interaction and facilitate the design of novel and innovative interfaces that can effectively serve users in complex and dynamic environments.

\begin{figure*}[tb]
 \centering 
 \includegraphics[width=\linewidth]{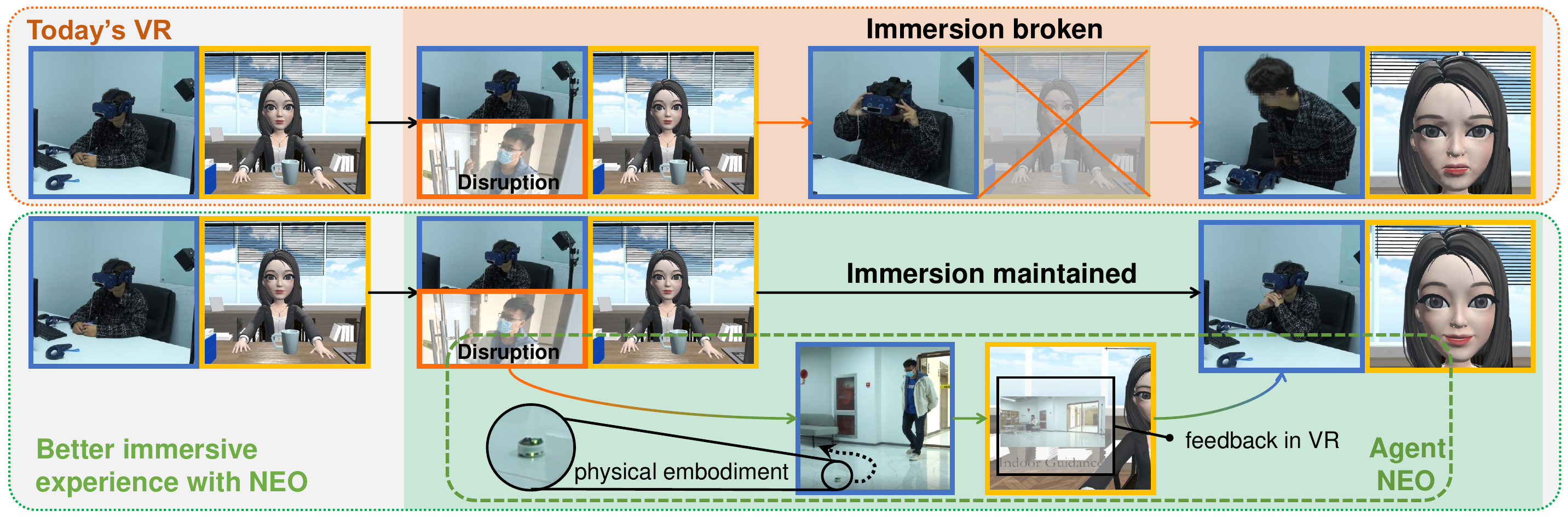}
 \caption{External stimuli and disruptions may break the immersion. We propose an intelligent agent NEO that could perceive user's demands, help them to handle the disruptions, and preserve their immersion in virtual environments.} 
 \label{fig:userpath}
\end{figure*}

We proposed an intelligent agent, the neighbor-environment observer (NEO), as shown in \autoref{fig:userpath}, which is designed to operate across the physical and virtual environments, accompanying human users in their daily activities. 
By parsing data from a non-invasive sensor system, NEO leverages a scalable decision module to generate real-time decisions that align with user needs without the need for explicit user commands.
We also provide a series of physical and virtual embodiments to take action across environments. 

We then conducted a user study to assess the usability of NEO.
The results demonstrate that NEO effectively reduces user workload resulting from disruptions during VR immersion, increasing user engagement.
These findings highlight the importance of developing agents that can seamlessly operate across physical and virtual environments, providing personalized and responsive support to users in dynamic and complex settings.

In summary, this paper makes four key contributions.
\begin{itemize}
\item We design a framework for neighbor-environment observation that leverages non-invasive sensors to gather information from both physical and virtual environments.
\item We propose a decision-making algorithm that can (1) generate decisions autonomously, (2) is compatible with different levels of sensor implementation, and (3) can learn user preferences interactively.
\item We propose a series of physical and virtual embodiments and a matching method so that they can collaborate to execute the generated decisions across both environments.
\item We develop a prototype system, NEO, as an artificial secretary in a smart home scenario, specifically a study room.
\end{itemize}

\section{Related Works}

\subsection{Environment Observation}  
Multiple sensors have been implemented in smart spaces to gain a better perception of the physical environment. 
Cameras and microphones are widely used across research and industrial fields. 
RGB cameras are used for 2D object detection, while depth sensors are usually used for 3D object detection and robotic navigation~\cite{haque2020illuminating}.
Thermal sensors are used to monitor temperature, which can also be used as safety alarms.
Dowdall et al.~\cite{dowdall2001millennium} proposed a system to support the life of senior people.
They used burglar alarm-style sensors to detect the states of doors and windows.
A custom-made body count sensor was used to detect the number of people in the environment.
Kidd's team~\cite{kidd1999aware} proposed the design of a system using small radio-frequency tags to track the movement of objects and show their locations.
Borelli et al.~\cite{borelli2019habitat} proposed an extensive digital platform for smart homes named HABITAT with the technology of the Internet of Things (IoT). 
They utilized radio frequency identification (RFID) readers and tags for localization.

\subsection{User Understanding}

Understanding human behavior and intention is a widely discussed topic. 
Thakur et al.~\cite{thakur2021ambient} used skeleton data collected by wearable sensors to analyze interactions of the aged group in smart homes. They infer emergencies by a rule-based reasoning diagram.

Attention, sometimes also denoted as engagement~\cite{szafir2012pay}, is another important criterion studied in smart-home-related research. 
Data is collected through different methods including questionnaires~\cite{oertel2020engagement}, RGB or depth cameras, temporal performances, interviews, observations, physiological sensors (e.g., electroencephalography), tracking sensors (e.g., motion, eye and laser tracking), speech and dialogue records, contextual and application record (e.g., game score), etc. Visual features such as head pose and eye gaze are widely used as a reference for attention understanding~\cite{danninger2008context, rudovic2019personalized, qiao2021review}. The algorithm may overlook evidence of distraction when only using one feature. For example, when the user stares at what he/she is supposed to focus on, it is difficult to detect the state of daze~\cite{qiao2021review}. Thus, some researchers turn to fusing modalities for better solutions~\cite{veras2022drivers, berrezueta2020smart}. Rudovic et al.~\cite{rudovic2019multi}, for example, fused data from an RGB camera, a microphone, and a smart watch to evaluate children's attention during child-robot interactions.

There has been a significant amount of work focused on the detection of interruptibility~\cite{korpipaa2004utilising, hudson2003predicting} using microphones~\cite{fogarty2004examining}, personal computer (PC) activity~\cite{horvitz2003learning}, and motion sensors~\cite{ho2005using}.

\subsection{Embodied Agent in Virtual Environments}
It is a rising topic that gives AI agents virtual embodiments and enables them to explore the virtual world.
There are virtual (or physical-virtual mixed) environments designed for agents to learn skills from human demonstration~\cite{cha2012virtual, li2017earthquake, schreuder2014validation}. Researchers also developed virtual simulators where AI agents learn new skills by interacting with the environments with their embodiments, without human demonstration. For example, CARLA~\cite{dosovitskiy2017carla}, AI2-THOR~\cite{kolve2017ai2}, Virtual Home~\cite{puig2018virtualhome}, iGibson~\cite{xia2020interactive,li2021igibson} are popular simulators.
Therefore, we believe that the embodiments are necessary to build an agent for immersive working companionship. 
We decided to extend the design of embodiments to the physical environment and propose a matching method enabling their collaboration.

\subsection{Agent in Smart Spaces}  

Some systems provide feedback by tuning the environment~\cite{zhao2017mediated, richer2020exploring} while others utilize avatars with digital cartoon faces~\cite{danninger2008context, berrezueta2020smart}, simplified robotic bodies~\cite{foster2017automatically, szafir2012pay}, or humanoid bodies~\cite{rudovic2019multi, hoppe2020human}.

Danninger et al.~\cite{danninger2008context} proposed a virtual secretary system in the office that mediated phone calls and visits by implementing an RGB camera and a microphone in the office.  
Zhao et al.~\cite{zhao2017mediated} and Richer et al.~\cite{richer2020exploring} designed a system that changed its appearance according to the users' physical states to help them keep concentrated or recover from stress. Cameras and wearable sensors were used to understand users' states while projectors, lighting systems, and sound systems were used to control the room's appearance.

\section{Neighbor-Environment Observation}
\subsection{Concept}
\textbf{Symmetrical Reality.}
\textit{Symmetrical Reality} (SR) is a framework that unifies concepts and interactions about the combination of the physical world and the virtual world~\cite{zhang2019symmetrical}.
The two worlds are topologically symmetrical.
The virtual world is composed of virtual agents and the virtual environment, similar to how the physical world is composed of humans and the environment. 
The virtual environment and virtual agents can evolve automatically, similar to how the physical environment and humans can evolve. SR defines a symmetrical topology between the physical and the virtual world, highlighting the importance of considering virtual agents and human users at an equal level.

\textbf{Neighbor-Environment Observation.}
In this article, we define the physical and the virtual environments as neighbor environments.
Following the SR paradigm, neighbor-environment observation is an approach to fuse information from these two types of environments.
As human users could interact in both environments, we suggest that only by observing the two environments simultaneously could the agent gain an overall knowledge of the ongoing situation.
Therefore, such methodology enables the agent to better parse the states of the users and the environments.
Here we consider the mapping between the physical environment and the virtual environment as a bijection.
Cases, where one physical environment is linked to multiple virtual environments, will be discussed in the discussion chapter.

\subsection{Requirements}
When users immerse themselves in a virtual environment through digital devices, their physical bodies continue to gather information from the physical environment around them. 
However, this information can often be detrimental to the immersive experience in virtual reality, as disruptions from the physical environment such as a ringing phone or unexpected guests can easily break the immersion. 
Users may be forced to leave the virtual environment to address them, which might interrupt an important task (e.g., a report at an international conference or a negotiation with the boss). 
We contend that intelligent agents in smart homes should be able to recognize such dilemmas and provide spontaneous assistance to alleviate these disruptions.
To achieve this, the agent must possess advanced cognitive abilities.
Here list the requirements:
\begin{enumerate}
    \item [Req-1:] Perception. The agent should be able to accurately perceive and understand the immediate situation, including the user states and the disruptions. (Related to \textit{RQ-1})
    \item [Req-2:] Reasoning. The agent should be able to reason about the user's availability to handle disruptions and determine whether it is appropriate to intervene. (Related to \textit{RQ-2})
    \item [Req-3:] Action. The agent should be capable of executing actions in both the virtual and the physical environments to minimize disruptions for the user. (Related to \textit{RQ-3})
    \item [Req-4:] Feedback. The agent should provide appropriate feedback about the occurrence of disruptions while minimizing negative impacts on their immersion. (Related to \textit{RQ-4})
\end{enumerate}

The four requirements are related to the four RQs accordingly. 

\textit{Req-1} demands the agent to take both the physical and the virtual environment into account. 
Thus, sensors should be implemented into the two environments.
Collected data should be aligned with a minimal time delay.
Moreover, it requires the sensors to be non-invasive so that they will not interfere with the user's primary tasks.
Data transfer should be wireless so that the agent can perceive a larger physical region.

\textit{Req-2} requires the agent to parse both the physical states and the virtual states of the user, as well as the environmental states.
The agent should be able to distinguish when is appropriate to help and when is unnecessary.
It requires an understanding of the user's goals and priorities.
As user preferences are various, the agent's mind should be compatible with different users and should be able to update itself.

\textit{Req-3} requires the agent to have the ability to execute actions in the two environments. 
Sometimes it needs to manipulate objects in the physical world to handle certain disruptions.
For example, when the battery is low, one should take off the VR HMD, find the charger, and plug it into the device.
Neither graphical notifications nor voice assistants could solve such problems.
Thus, we present a series of embodiments including physical ones like desktop robots and virtual ones such as notification user interfaces (UIs).

\textit{Req-4} demands the agent to understand the user's preferences and to provide clear and informative feedback.
The feedback should be straightforward to alert the immersed user that something has happened and what has the agent done for the user.
It should also be lightweight to avoid affecting others who are sharing the physical or virtual environment with the user.
In addition, the agent should have a port to collect users' feedback and update itself accordingly.

\section{Method}

\subsection{Design Overview}
In this section, we propose a framework for intelligent agents to observe, understand and cooperate with humans when they immerse in a virtual environment.
The framework is designed for single-user use.
It is constructed by three modules: the perception module, the decision module, and the action module.
Each module is designed with a specific purpose, utilizing distinct methods. 
We will provide a detailed explanation of these module individually.

\begin{figure}[tb]
 \centering 
 \includegraphics[width=\columnwidth]{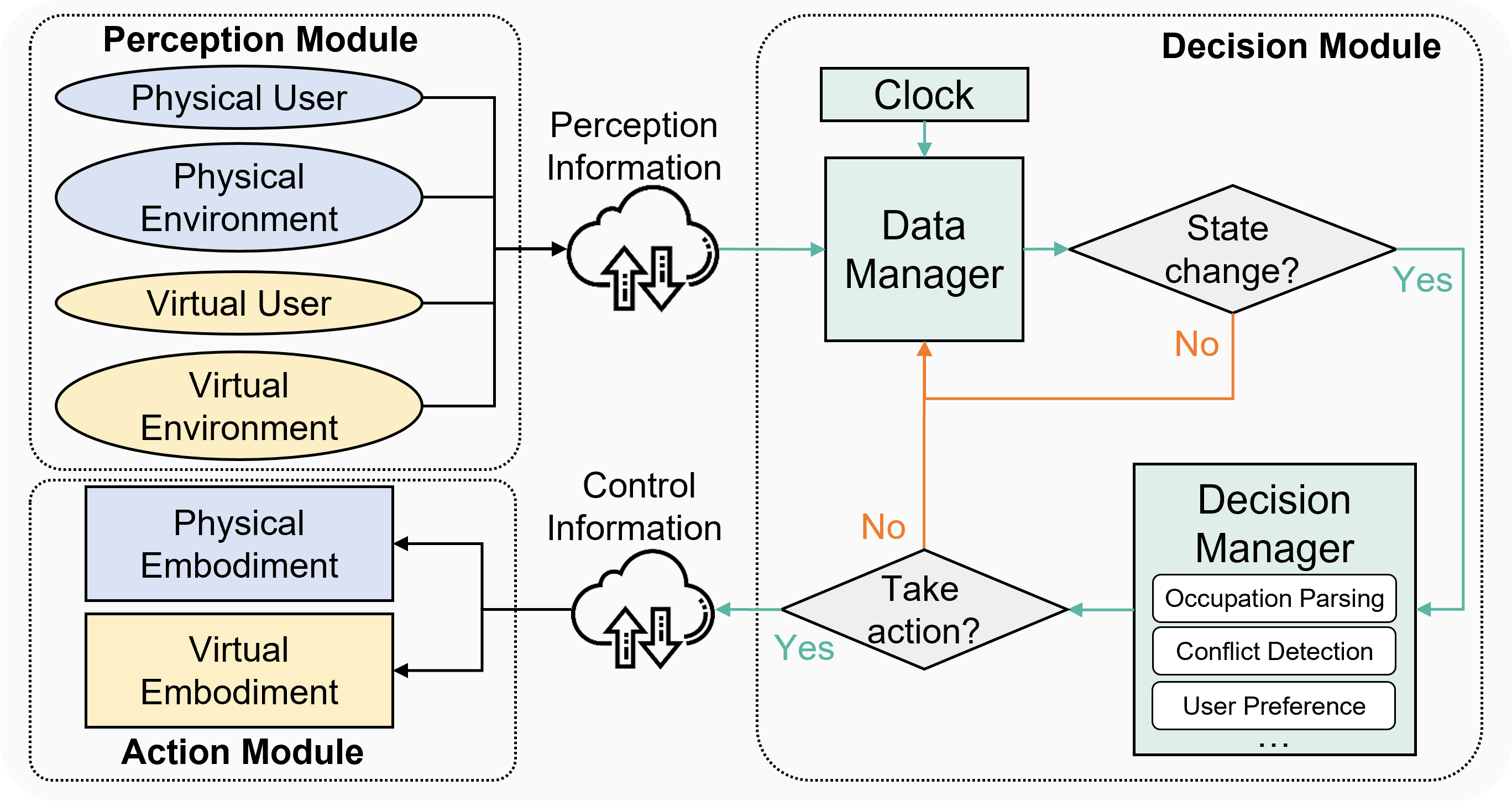}
 \caption{Information and decision transfer process in our framework. Joint observation information is transmitted to a data manager via WiFi. The data manager activates the decision manager, from which the decision is transmitted to the two types of embodiment.}
 \label{fig:overview_dataflow}
\end{figure}

\autoref{fig:overview_dataflow} shows an overview of how the information transfers between the three modules.
Our framework is developed with the setting of common immersive work scenarios, utilizing a personal computer (PC) and a VR HMD.
This allows quick evaluation of effectiveness with minimal use of customized sensors and devices. 
To increase portability, we employ wireless methods for data transfer.

\subsection{Joint Observation}
This method is focused on enabling the agent to access both physical and virtual environments, expanding the current capabilities of smart home agents, which primarily perceive the physical world. 
Physical sensors act as tentacles to gather information from the physical environment, while virtual data ports serve as tentacles to the virtual environments.

The collected information is grouped into four categories: (1) physical environmental information (PE); (2) physical user information (PU); (3) virtual environmental information (VE); and (4) virtual user information (VU). 
The sources included in this work are summarized in \autoref{tab:sources}.

\textbf{Physical environmental information. }
We add two additional sensors (two RGB cameras) to capture visual information in the physical environment.
From this data, the agent detects events and disruptions that occur in the physical environment.
The cameras' positions define the agent's perception area.
In the prototype system, as shown in \autoref{fig:teaser}, one camera monitors the user's working area near the desk, while the other watches the door for visitors. 
The data from each camera is parsed independently. 
Our parsing algorithm does not require binocular-camera reconstruction, making it easy to change the number of cameras without calibration.

\textbf{Physical user information. }
Visual and auditory information is utilized to interpret the user's states. 
An RGB camera is located beside the desk.
The system classifies user poses into nine categories: using a device, using a keyboard, using a mouse, writing, reading, using a mobile device, resting, drinking, and eating.
A microphone is used to detect whether the user is speaking.
The system also infers three physiological needs by the time interval between the current time and the last detection of the corresponding action: thirst (activated after 30 minutes), hunger (activated after 180 minutes), and fatigue (activated after 2 hours of continuous work as prolonged periods may harm health~\cite{young2016sedentary}, or after 20 minutes of VR usage considering visual fatigue~\cite{lee2021visual}).
Users can adjust these thresholds.

\textbf{Virtual environmental information. }
The activated window on the PC desktop and the activated scene in VR are recorded.
We define three kinds of mappings for PC windows: work (online meeting apps, browsers, Microsoft Office Word), entertainment (PC games, media players), and others.
Similarly, we define three kinds of mappings for VR scenes: work (online meeting scenes), entertainment (VR games, painting Apps), and others.
A thread is activated to monitor the battery levels and data from the controllers.

\textbf{Virtual user information. }
We monitor four user activities: mouse usage, keyboard usage, VR HMD usage, and VR controller interactions. 
Mouse and keyboard activities pertain to interactions through the PC. 
The VR HMD position indicates the user's location in the virtual environment.
With calibration, it also shows their physical location.
The HMD's orientation shows the user's gaze direction. 
Additionally, input signals from the VR controllers are utilized to recognize the user's options, indicating whether the user is occupied or not.

\begin{table}
  \caption{Information sources in joint observation}
  \label{tab:sources}
  \centering
  \resizebox*{0.7\linewidth}{!}{
  \begin{tabular}{rl}
    \toprule
    Information Category & Type of Sources\footnotemark[1]\\
    \midrule
    \multirow{2}*{Physical Environment (PE) } & \multirow{2}*{Visual information \textit{(RGB camera)}. } \\
                                           ~ & ~ \\
    \midrule
    \multirow{2}*{Physical User (PU)  }      & Visual information \textit{(RGB camera)};   \\
                                           ~ & Auditory information \textit{(Microphone)}.   \\
    \midrule
    \multirow{2}*{Virtual Environment (VE) } & Activated windows on PC; \\
                                           ~ & Activated VR scenes. \\
    \midrule
    \multirow{2}*{Virtual User (VU)   }      & Activity of mouse and keyboard;  \\
                                           ~ & Activity of VR HMD \& controllers.  \\
    \bottomrule
    \end{tabular}
    }\\
    \footnotesize{$^1$ Italics indicate that additional sensors are utilized to capture such information.} 
\end{table}

\subsection{Scalable Decision Making}
\subsubsection{Decision Module}

The decision module is designed based on the information type from the perception module.
It is initialized by an And-Or Graph (AOG)~\cite{zhu2007stochastic} with a probabilistic model.
When receiving reports from the sensors, the decision module generates a parse tree by pruning the AOGs and categorizes possible user occupation states into two channels: 
\begin{enumerate} 
    \item \textbf{input channel:} hands occupation, speaking occupation.
    \item \textbf{output channel:} visual occupation, auditory occupation.
\end{enumerate}

\begin{figure*}[tb]
 \centering 
 \includegraphics[width=\linewidth]{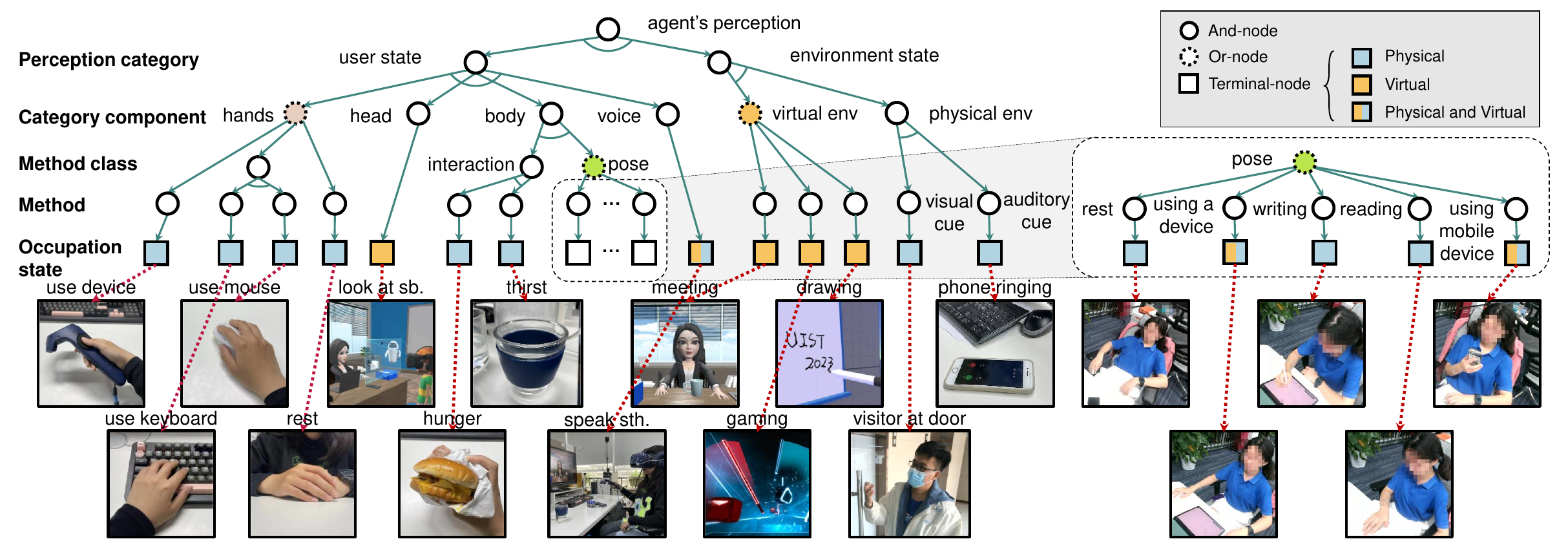}
 \caption{The And-Or Graph of the joint observation. The perception of the agent is categorized into two sets: user state and environment state. The user state node consists of four components, representing occupation methods, which are further divided based on method classes and specific methods. The environment state node consists of two components: the virtual part and the physical part. Colors of the Terminal-nodes denote whether the nodes belong to the physical environment or not.}
 \label{fig:AOG}
\end{figure*}

\textbf{Joint parsing.}
To give a better vision of this model, we utilize the case where the user works from home using a VR HMD and PC.
The AOG, shown in \autoref{fig:AOG}, consists of two halves: the left half represents joint user observations ($G_u=(V_u, E_u)$), and the right half represents joint environment observations ($G_e=(V_e, E_e)$).
The AOG contains And-nodes, Or-nodes, and Terminal-nodes~\cite{zhu2007stochastic}.
Different layers represent different levels of observation.
Terminal-nodes represent the occupation states to be observed.

\textbf{Occupation inference.} 
The decision module utilizes the parse tree generated from the AOG to infer potential occupations caused by disruptions and check for conflicts with existing occupations.
The Terminal-nodes of the parse tree vote for the overall occupation.
We introduce a hyperparameter to alter the weight between positive votes and negative votes.
The agent infers the overall occupation state according to the percentage of positive votes among all votes.

\textbf{Probabilistic model.}
The probability of taking actions at time $t$ is denoted as \autoref{eq:prob}.
\begin{equation}
  P(A_t, C_t, pt_t) = P(A_t|C_t, pt_t)P(C_t|pt_t)P(pt_t)\label{eq:prob}
\end{equation}
where $A_t$ denotes the decided action at time $t$, $C_t$ denotes the conflict at time $t$, and $pt_t$ denotes the parse tree at time $t$.

\subsubsection{Personalization}

\begin{figure}[tb]
 \centering 
 \includegraphics[width=0.7\columnwidth]{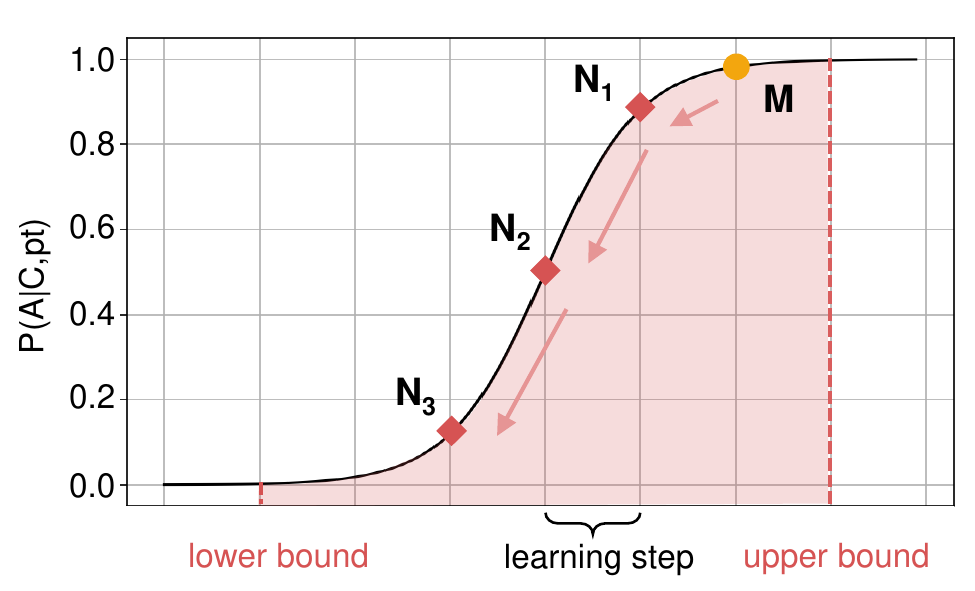}
 \caption{The agent adjusts the conditional probability along the sigmoid curve according to the user's feedback. It stops when reaching the upper or lower bound.}
 \label{fig:learning}
\end{figure}

We propose a personalization module to adjust the probabilities of taking actions so that the agent learns the personal preferences of different users.
We utilize a sigmoid function denoted as \autoref{eq:sigmoid} when tackling $P(A_t|C_t, pt_t)$.
\begin{equation}
  sigmoid(x) = \frac{1}{1+e^{-z}}\label{eq:sigmoid}
\end{equation}
For each $P(A|C, pt)$, we define an initial probability of taking action.
After each decision the agent made, the user gives feedback to the agent for their action.
According to the feedback, the conditional probability of taking action $A_t$ oscillates between 0 and 1 along the sigmoid function.
\autoref{fig:learning} shows an example.
Point M shows the initial conditional probability. 
If the agent receives negative feedback, the probability takes one step left to reach point $N_1$ and takes $sigmoid(N_1)$ as the new conditional probability, and vice versa. 
If the user gives negative feedback again, the probability goes to point $N_2$ to update $P(A|C, pt)$.

\subsection{Joint Action}
To take action and offer assistance in the two environments, the agent needs different kinds of embodiment.
We define the actions into four categories.
\textbf{Graphical actions} in the virtual environment are shown as UI notifications, including (1) a reminder of taking a break; (2) a video stream of the visitor at the door; (3) a reminder of the devices' low battery state.
\textbf{Virtual embodied actions} are designed to provide visual cues in the virtual environment. A robot character is shown if NEO starts monitoring. When bringing objects for the user, NEO renders their virtual counterparts for the user to grab the delivered gadgets without removing VR HMD.
\textbf{Physical embodied actions} are designed for the agent to interact with the physical environment. 
The agent brings desktop-level objects and receives visitors.
Details are introduced in the next chapter.
\textbf{Auditory actions} play an auxiliary role in receiving visitors.

\begin{figure}[tb]
 \centering 
 \includegraphics[width=0.7\columnwidth]{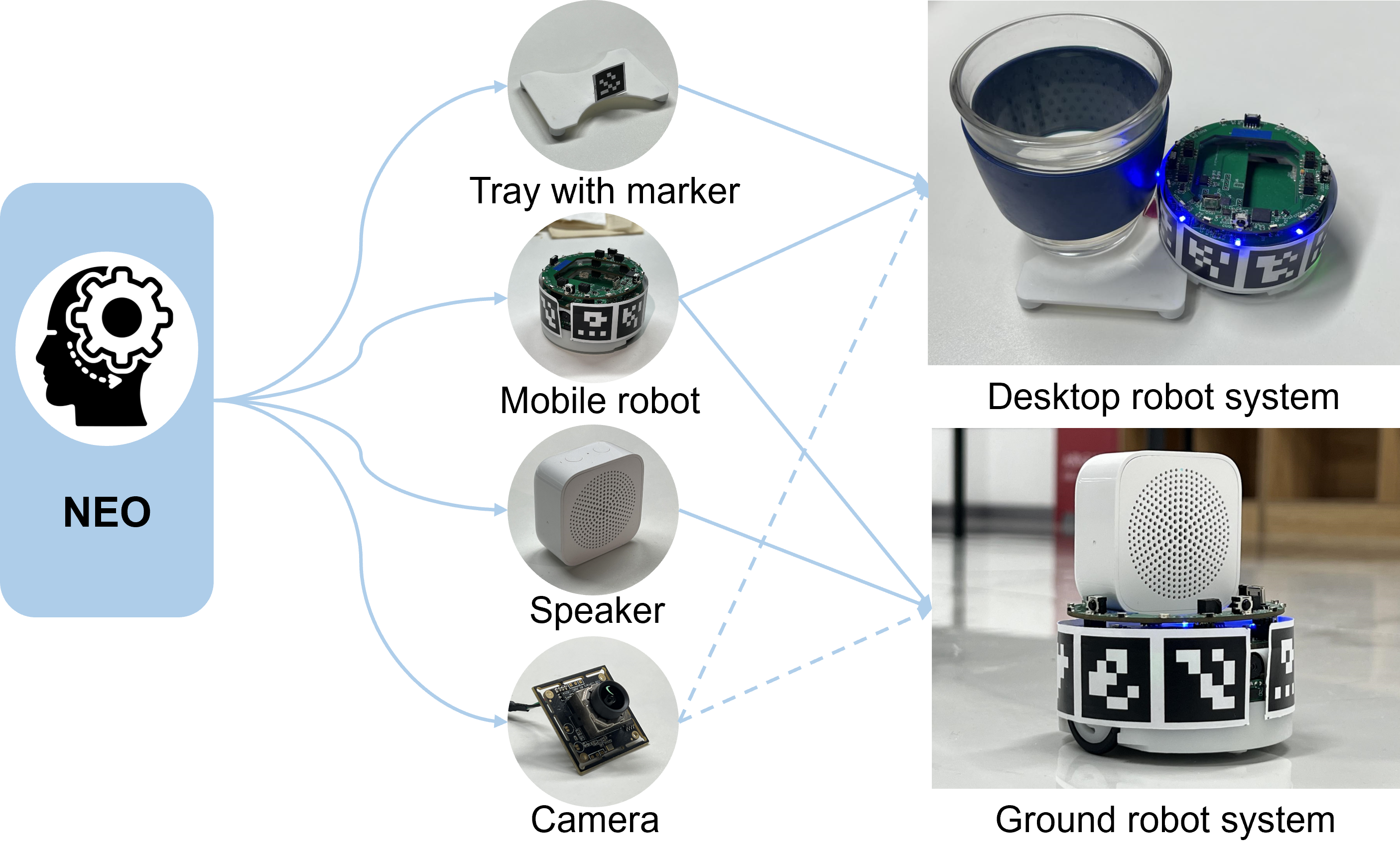}
 \caption{Hardware implementation of NEO. The mobile robots serve as movable physical embodiments of NEO. Together with trays and the camera, they construct the desktop robot system that provides object transportation. The mobile robot connected with a speaker forms the ground robot system that could help the users to receive visitors.}
 \label{fig:hardware}
\end{figure}

\section{Implementation}
Following these methods, we develop a prototype system, NEO.

\subsection{Software Overview}

\textbf{Algorithm implementation. }
To simplify the system, similar algorithms are used to process data from similar sensor types (See \autoref{tab:sources} for sensors of NEO). 
For instance, images from RGB cameras are processed by similar algorithms developed based on CVZone~\cite{cvzone}, OpenCV~\cite{bradski2000opencv} and MediaPipe~\cite{lugaresi2019mediapipe}. 
Algorithms for cameras of the physical environment detect and crop region that includes faces in the image to detect visitors and inform the user.
Algorithms for cameras of the physical user estimate the user's skeleton and detect objects near the user to categorize the pose.
Virtual activities data of the user are smoothed by a filter before passing to the decision module.
Other data is transferred to the decision module directly.

\textbf{Interface implementation. }
In order to facilitate NEO's graphical actions, we design a series of interfaces.
To show the design, we use the scenario of a one-on-one VR meeting as an example.
As shown in \autoref{fig:teaser}, NEO renders a robot character at startup to represent its running state to the user.
NEO renders a virtual counterpart in the virtual environment when bringing an object to the user physically. 
By calibrating the position of NEO’s physical embodiments, we align the position of physical objects and their virtual counterparts. 
For example, if NEO brings a cell phone to the user, a virtual phone is rendered in the VR scene, moving in sync with the physical phone.
All virtual interface is only visible to users. Other individuals in the VR scene are unable to see the virtual objects and interfaces.

\textbf{Data communication and management. }
A data manager module is developed to align the heterogeneous input data by a clock wave (T = 20 ms), as shown in ~\autoref{fig:overview_dataflow}.
The sensor states are checked at the clock-rising edge. 
If a sensor does not send data in the last clock period, its state will be set to \textit{inactive}. To improve efficiency, the data manager module passes current states to the decision module only when a state change occurs.

\begin{figure*}[tb]
 \centering 
 \includegraphics[width=\linewidth]{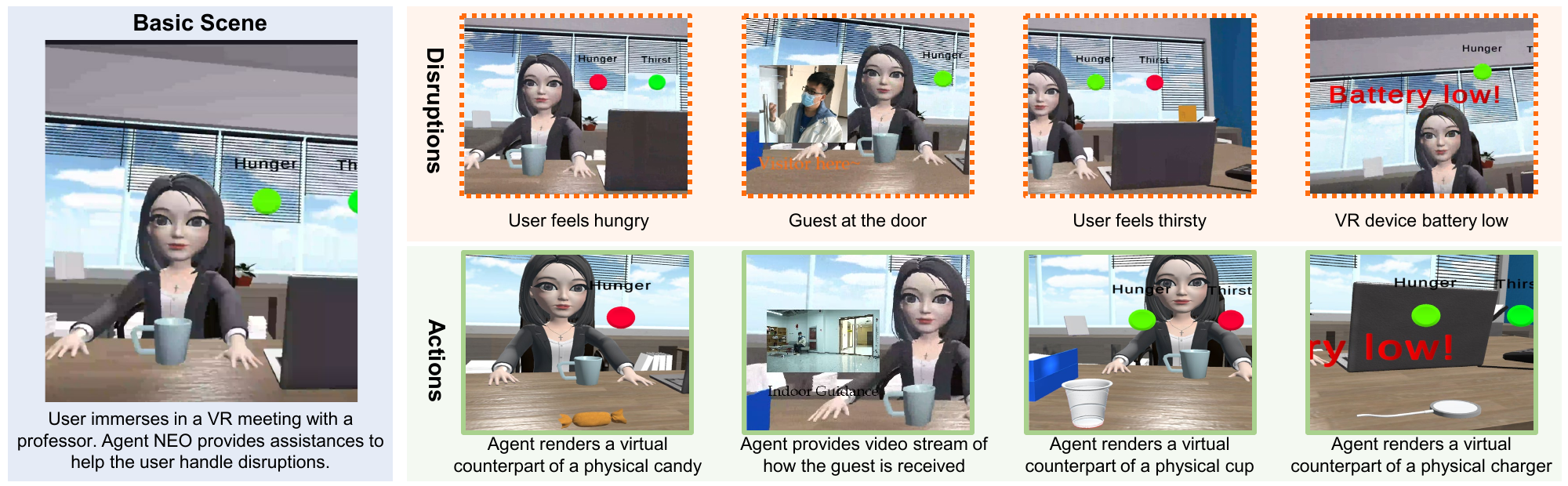}
 \caption{VR implementation of NEO. It visualizes how the disruptions and actions are rendered in VR.}
 \label{fig:UI}
\end{figure*}

\subsection{Hardware Overview}
Hardware implementation is shown in \autoref{fig:hardware}.

\textbf{Personal computer. }
The core computation of NEO is accomplished by the existing personal computer of the user.
In this work, we implement NEO on a PC with an AMD Ryzen 9 5950X CPU and an NVIDIA RTX 3070 GPU. 
It is also used for rendering virtual environments for the experiment.
Two 1080p RGB cameras are connected to the PC to capture visual information. 

\textbf{VR related devices. }
An HTC Vive HMD and two controllers are used. 
The HMD displays the virtual environment, while the controllers allow for interaction. 
They also provide VU sensors for NEO: the HMD provides the user's orientation and the built-in microphone collects auditory cues.
A Leap Motion device is mounted in the front of the Vive HMD to reconstruct the skeleton of the user's hands for object grasping.

\textbf{Movable embodiment. }
A desktop robot with a diameter of 7 $cm$ is developed as the movable embodiment. The robot is equipped with two differential-driven wheels and two reduction stepper motors, by which it can move freely on the tabletop and locate itself by odometry. The linear speed of the robot can reach 10 $cm/s$ with 3~$cm$ diameter wheels, and the pull-out torque can reach 0.4 $kgf/cm$. With a wireless microcontroller unit (ESP32PICOV302) the robot is able to receive the message either from WiFi or from Bluetooth Low Energy (BLE) devices. 
Trays with markers and universal wheels (4 wheels for 1 tray) are 3D printed to support object transportation.
The desktop robot could identify markers on the tray to locate the target objects on it.
To bring objects to the user, the desktop robot moves to the marker, and pushes the tray toward the user.

\subsection{VR-side Implementation}

To help readers better understand our framework, we demonstrate it in a VR experience of a one-on-one meeting.
\autoref{fig:UI} shows the basic scene of the implementation, some examples of the disruptions, and the corresponding actions provided by NEO.
During the meeting, the VR users discuss a research topic with the avatar of a professor.
The professor introduces her point of view in the meeting and brings up several open questions.
Different types of disruptions, either from the physical environment or the virtual environment, are rendered during the meeting.
The VR users are asked to concentrate on the meeting as much as possible while handling these disruptions.

\section{Experiment}

\begin{table*}[tb]
    \caption{Dimensions and cases in the test set.}
    \label{tab:testset}
    \small

    \resizebox*{\linewidth}{!}{
    \begin{tabular}{cl|cl|cl}
    \toprule
    \multicolumn{2}{c|}{\textbf{Activated Sensor Group}} & \multicolumn{2}{c|}{\textbf{Occupation}} & \multicolumn{2}{c}{\textbf{Disruption}}\\
    \midrule
    \textbf{ID} & \textbf{Case} & \textbf{ID} & \textbf{Case} & \textbf{ID} & \textbf{Case}\\
    \midrule
    S1 & PE only & O1 & Have a one-on-one discussion in VR & D1 & Visitor knocks at the door \\
    S2 & VU only & O2 & Play a game in VR & D2 & User feels thirsty \\
    S3 & PU only & O3 & Paint in VR with controllers & D3 & User keeps using digital device over 20 minutes \\
    S4 & VU-VE & O4 & Have a multi-person meeting in VR & D4 & Phone rings \\
    S5 & PE-VU & O5 & Watch a movie in VR & D5 & VR device notifies battery low \\
    S6 & PU-PE & O6 & Have an online interview using PC & D6 & User knocks over a glass of water \\
    S7 & PU-VU-VE & O7 & Play game on PC & & \\
    S8 & PU-PE-VU & O8 & Write an article on PC & & \\
    S9 & PU-PE-VE & O9 & Have an online lesson on PC & & \\
    S10 & PU-PE-VU-VE & O10 & Watch a movie on PC & & \\
    & & O11 & Use a mobile device (smartphone or tablet) & & \\
    & & O12 & Read a book & & \\
    & & O13 & Have a rest & & \\
    \bottomrule
    \end{tabular}
    }
\end{table*}

We launched two technical experiments to evaluate the ability of NEO from two perspectives: (1) evaluation \#1: the influence of different activated sensor groups; (2) evaluation \#2: NEO's ability to learn the user's preference.

\subsection{System Evaluation \#1}
\subsubsection{Experiment design}
\textbf{Evaluation goal.} 
Measure the extent of NEO's capability when activating different sensor groups, related to the \textit{Req-1} and \textit{Req-2}.

\textbf{Apparatus.}
We prepared a toy test set comprising different online working or entertaining scenarios with two dimensions: occupation, and disruption.
There are 13 cases in the occupation set (O01-O13) and 6 cases in the disruption set (D1-D6).
O13 was included as it is common in daily life. 
Testing this case will make the system evaluation more comprehensive.
Sensors were categorized into four classes according to their information source type listed in \autoref{tab:sources}: physical environment (PE), physical user (PU), virtual environment (VE), and virtual user (VU).
To make the result clearer, we chose 10 typical sensor combinations (S1-S10).
\autoref{tab:testset} shows the details.

\textbf{Task.}
We explored the system's ability by activating different sensor groups and testing the decision accuracy of NEO.
The accuracy here is defined as \autoref{eq:acc}:
\begin{equation}
  Acc = \frac{\sum_{d \in D}correct(d)}{N_{decision}}\label{eq:acc}
\end{equation}
where $N_{decision}$ is the total number of decisions in a certain case, D is the set of decisions from NEO in this case, and $correct(d)$ is defined as:
\begin{equation}
\label{eq:correct}
correct(d) =\left\{
    \begin{array}{lcl}
    1 & , & d = d_{gt}, \\
    0 & , & else.
    \end{array}
\right.
\end{equation}

\subsubsection{Result}
\begin{figure*}[tb]
 \centering 
 \includegraphics[width=\linewidth]{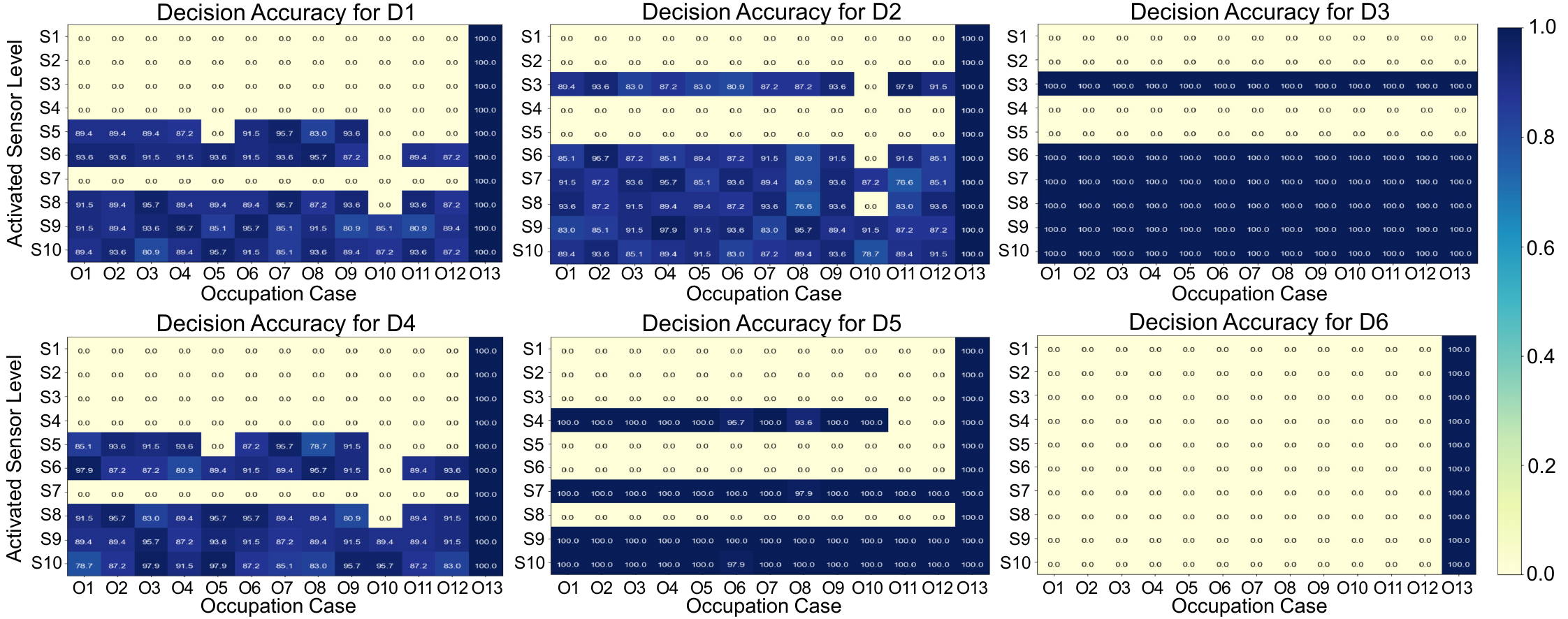}
 \caption{Result of system experiment. The six sub-graphs visualize the decision accuracy of NEO in the occurrence of the six disruption cases (D1-D6) respectively. The x-axis of each graph denotes the ID of the tested occupation case (O1-O13). The y-axises denote the ID of the activated sensor group (S1-S10). All details of the three groups of variables are shown in \autoref{tab:testset}. Numbers in the matrix refer to accuracy in terms of percentage. The percent symbol is hidden for a clearer view.}
 \label{fig:sysexp1} 
\end{figure*}

The result is shown in \autoref{fig:sysexp1}.
We visualized a total of 78 different cases with 10 different combinations of sensors.
Overall, the results show an increasing trend in accuracy when activating more sensors.
The performance on O13 reaches 100\%, showing that NEO knows when not to help. 
The level S10 that has access to all four observation classes reaches the highest performance.
S10 is able to handle most of the cases except for D6.
All sensor level shows zero accuracy in this case (except for O13).
It is because the current version of NEO fails to parse the disruption of ``knock over a glass of water'' so its decision remains ``no action''.
Disruptions like D6 are denoted as hard problems.
However, such shortcomings of NEO could be reduced in future work by extending its parsing algorithm.

For disruption case 1 (D1), we noticed that S7 failed to react to all the disruptions.
This was caused by the missing perception of the physical environment.
Information of D1, a visitor at the door, can only be observed in the physical environment.
Losing connection to PE sensors caused this fatal problem.
Similar phenomena were observed in D4 (S7), D5 (S8), etc. 
These phenomena support the theory of joint observation: losing connection to either the physical environment or the virtual environment will lead to a drop in performance.

O10 is a relatively demanding task where NEO's performance is sensitive to the information of VE.
Clues from PU are similar to O13.
Only by observing the virtual environment, namely the active window on the PC, can NEO understand the true state.
This also supports the joint observation theory.

\subsection{System Evaluation \#2}

\subsubsection{Experiment design}
\textbf{Evaluation goal.} 
Measure the capability of NEO to learn user preferences while testing whether the system fulfills \textit{Req-2}.

\textbf{Apparatus.}
We chose a typical case (O1+D1) from the test set to evaluate the ability of personalization.
The occupation case was set to always be ``have a one-on-one discussion in VR'' while the disruption was always ``a visitor knocks at the door''.
We put three simulated users in the simulated environment with different preferences about NEO's actions, which is shown in \autoref{tab:avatar}.
User A always gets annoyed if NEO goes and receives the visitor.
User B gets annoyed in the first 3 rounds, shows neutral attitudes in the middle 4 rounds, and reacts satisfied in the last 3 rounds.
User C is always satisfied with NEO's help.

\textbf{Task.}
NEO was allowed to interact with the three users 10 times to learn their preferences.

\begin{table}[b!]
    \caption{Personality of the simulated users.} 
    \label{tab:avatar}
	\centering%
    \begin{tabular*}{0.7\textwidth}{%
        r%
        *{7}{l}%
	*{2}{c}%
    }
    \toprule
    \textbf{User} & \textbf{Personality}\\
    \midrule
    User A & Always annoyed if NEO goes and receives the visitor. \\
    User B & Annoyed at first, neutral in the middle, satisfied at last. \\
    User C & Always satisfied if NEO goes and receives the visitor. \\
    \bottomrule
    \end{tabular*}
\end{table}

\subsubsection{Result}
\begin{figure}[tb]
 \centering 
 \includegraphics[width=0.7\columnwidth]{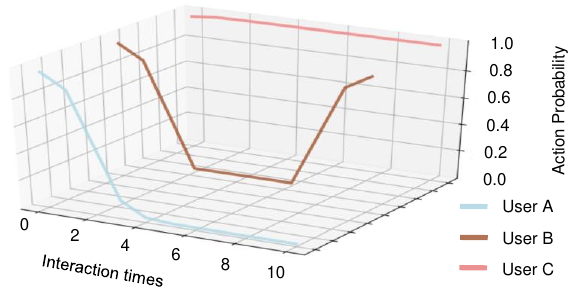}
 \caption{Result of learning experiment with three users.}
 \label{fig:sysexp2}
\end{figure}

The change in the action probability during the 10-round interactions is visualized in \autoref{fig:sysexp2}.
In the case of User A, the probability of receiving the visitor drops fast after the first three interactions.
In the case of User B, a similar drop is seen in the first two rounds.
The probability raises back after User B changes its mind at the last three interactions.
In the case of User C, the action probability remains high as User C likes NEO to receive visitors.
The result visualizes the learning procedure of NEO and demonstrates the ability of NEO to learn different preferences.

\subsection{User Study}

\subsubsection{Goal and Hypothesis.}
\textbf{Evaluation goal.} 
The goal of this user study is to validate whether our method can autonomously generate decisions when disruptions occur and reduce users' distraction through the embodiments.
Moreover, we want to evaluate the usability of NEO.
We invited the participants to come to our lab and experience a VR meeting.
During the meeting, we rendered multiple disruptions around the participants to record their reactions and feedback.

\textbf{Hypothesis.}
While the two system evaluations address \textit{Req-1} and \textit{Req-2}, the following hypotheses pertain to \textit{Req-3} and \textit{Req-4}.

\textbf{(H1) NEO could enhance users' sense of engagement during a VR meeting in the presence of external disruptions.}
We hypothesize that NEO's assistance in neighboring environments thereby increases user engagement when disruptions occur.
\textbf{(H2) NEO could alleviate users' workload in managing distracting events.} 
We hypothesize that when NEO shares a portion of the workload to disruptions, that of users is reduced.
\textbf{(H3) The personalization module would improve the overall user experience of the system.}
Although the action rate is designed based on occupation conflict, we acknowledge that people have varying preferences regarding NEO's assistance. 
Therefore, we postulate that user experience will be further enhanced after personalization.

\subsubsection{Participants}

Twelve participants (P01-P12, 5 Females and 7 Males) aged between 21 to 31 were recruited (M=24.92, SD=3.20). 
About 83.3\% of the participants have used VR devices and 66.7\% work online over eight times per month, which is considered relatively frequent. 
As compensation for their effort, participants received a small gift.
All participants attended in person and reached the end of the user study.

\subsubsection{Procedure}
The whole procedure lasted for 30 minutes. 
An investigator helped the participants to wear the VR HMD and took notes during the study. 
After reporting demographic information, participants were required to go through three phases of tasks: (1) the Control phase, (2) the Original NEO (O-NEO) phase, (3) the Personalized NEO (P-NEO) phase. 
If participants get familiar with the procedure, the ratings of engagement might naturally increase and the workload might drop.
To exclude such influence, we disrupted the order of the phases. 

\begin{figure*}[tb]
 \centering 
 \includegraphics[width=\linewidth]{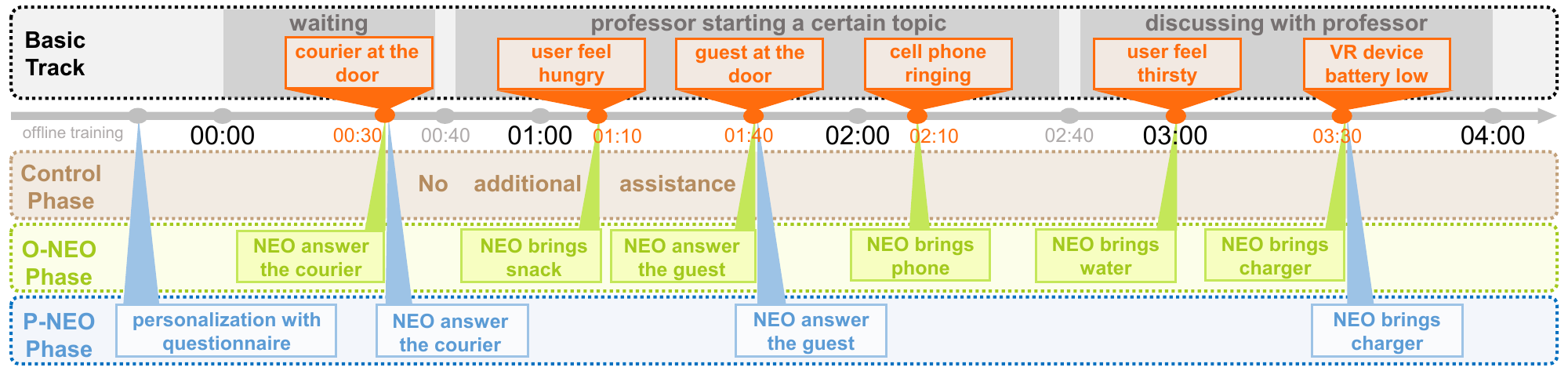}
 \caption{Procedure of the user study. The three phases shared the same basic track: the participant was asked to have a discussion with a professor. Several disruptions were rendered and the participant was required to handle them. Assistance in the P-NEO phase was different according to their preference. What is shown in this figure is an example of a possible case.}  
 \label{fig:user_study_pro}
\end{figure*}

The procedure is visualized in \autoref{fig:user_study_pro}.
Participants were required to focus on a discussion with a professor through a VR meeting. 
Six disruptions occurred in turn during the meeting.
The professor discussed different topics in different phases to exclude the influence of familiarity.
The three phases shared this same basic track:

    [00:00] Beginning of the experiment;
    
    [00:30] Courier visiting;
    
    [00:40] Professor starting a certain topic;
    
    [01:10] User feeling hungry;
    
    [01:40] Guest arriving;
    
    [02:10] Cell phone ringing;
    
    [02:40] Professor raising a question about the topic;
    
    [03:00] User feeling thirsty;
    
    [03:30] VR device battery low.

We rendered an alternative notice in the VR scene for physiological need, as shown in \autoref{fig:UI}.
The participants were asked to pretend that they were hungry or thirsty when the corresponding buttons turned from green to red.
We rendered the disruption of ``battery low'' similarly.
Note that both the alternative notice and the virtual interfaces were exclusively visible to the participants.

\begin{figure}[tb]
 \centering 
 \includegraphics[width=0.6\columnwidth]{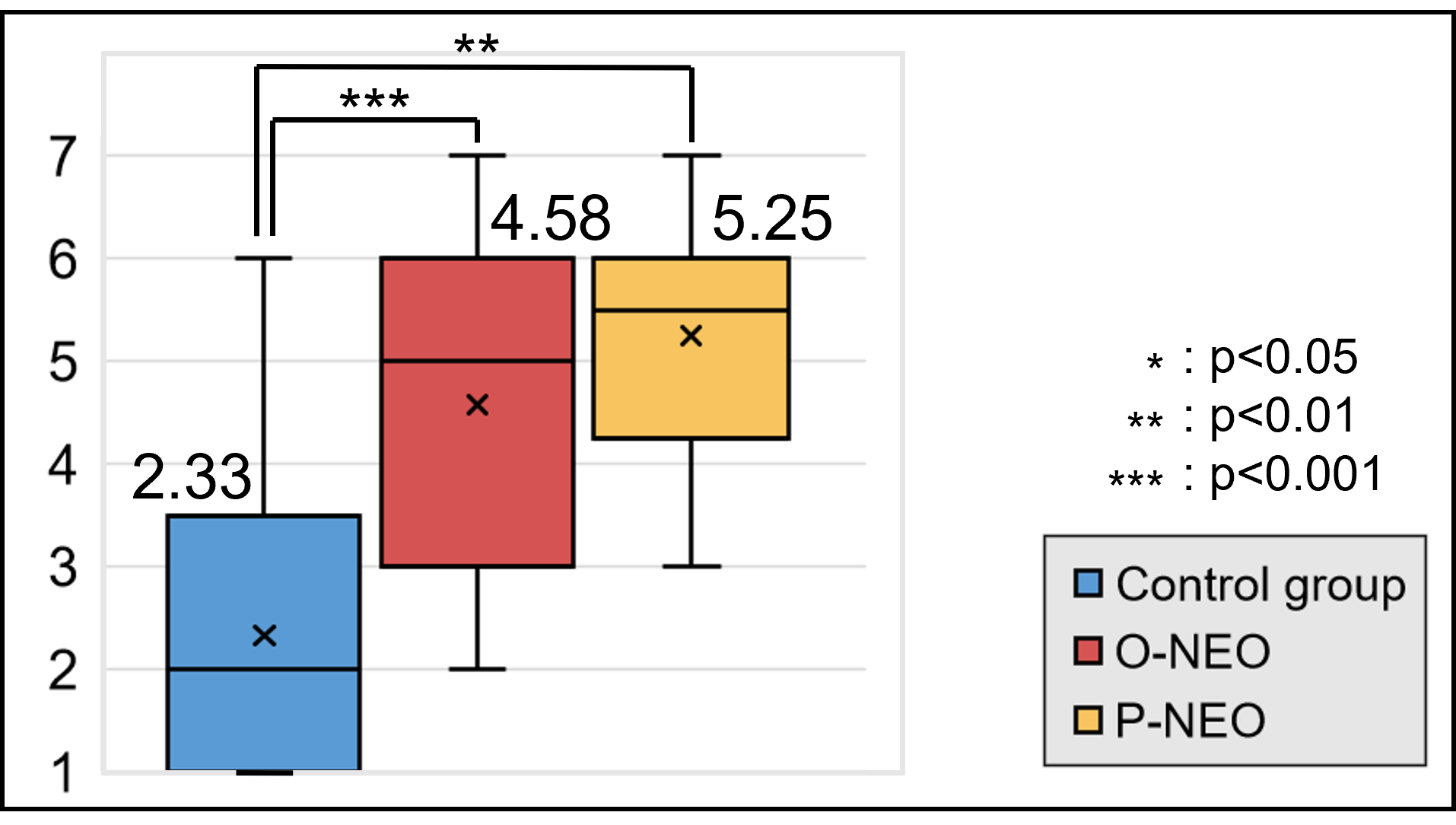}
 \caption{Ratings for engagement of VR meeting.}
 \label{fig:engagement}
\end{figure}

Each phase had its unique parts as listed below:

\textbf{Control phase:} No additional assistance was provided.

\textbf{O-NEO phase:} In this phase, O-NEO helped the participants to deal with various disturbances regardless of their preferences.

\textbf{P-NEO phase:} Before the start of this phase, all participants were asked to complete a personalized questionnaire for NEO to learn their preferences. During this phase, P-NEO took action selectively.

At the end of each phase, participants filled in a post-study questionnaire.
The post-study questionnaire consists of four parts: (1) a questionnaire asking participants to assess their engagement during the meeting~\cite{celiktutan2017multimodal, ben2019fly} (related to \textbf{H1}); (2) two NASA Task Load Index (NASA-TLX) questionnaires~\cite{NASATLXhart1988development} to evaluate the overall workload of the whole phase and that of handling the disruptions (related to \textbf{H2}); (3) a short version of the user experience questionnaire (UEQ-S)~\cite{UEQschrepp2017design} to quantify the user experience of O-NEO and P-NEO (related to \textbf{H3}); (4) a questionnaire to assess participants' satisfaction about O-NEO and P-NEO (related to \textbf{H3}).
For consistency, all measures are on a 7-point Likert (1=very low, 7=very high).

A semi-structured interview was launched at last.

\begin{figure*}[tb]
 \centering 
 \includegraphics[width=\linewidth]{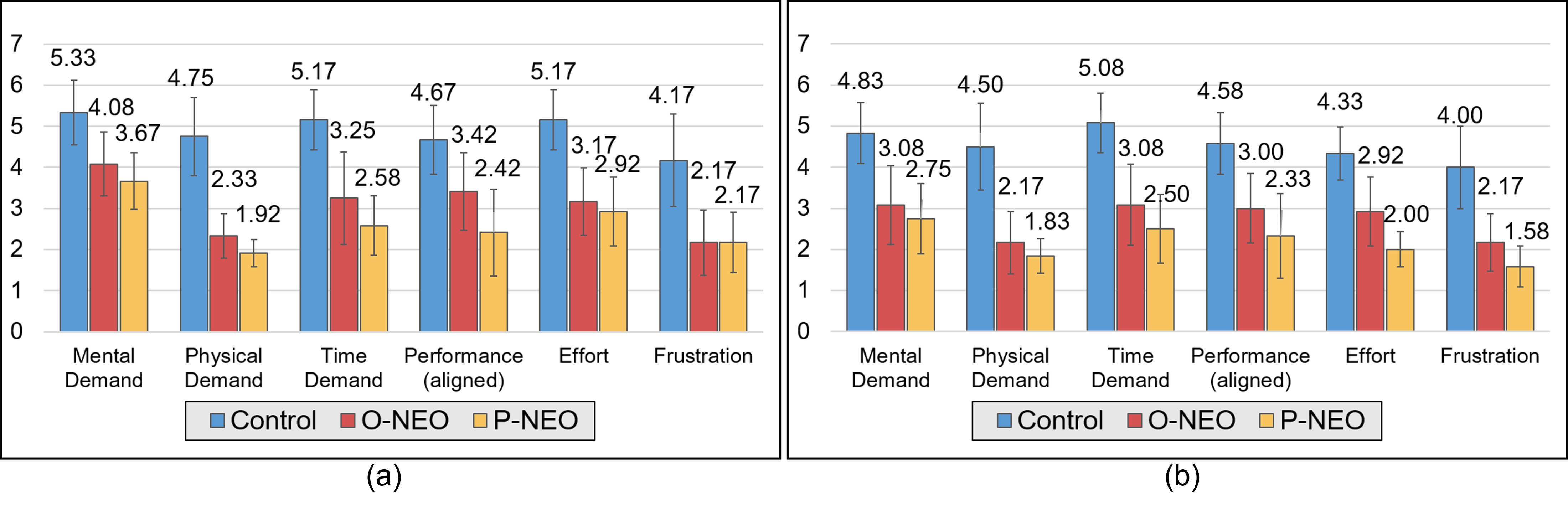}
 \caption{NASA-TLX results. (a) The overall workload of the whole task (including both discussing with the professor and handling disruptions ). (b) The workload of handling the disruptions. Note that the points of performance in this figure are aligned so that all questions share the same trend: ``the lower the point is, the better''. ($shown point = 7 - original point$)}
 \label{fig:NASA_TLX}
\end{figure*}

\subsubsection{Result}
We utilized paired t-tests for significance analysis between every two phases if the data fit a normal distribution.
We utilized the Wilcoxon signed rank tests when the normality assumption was violated.

\textbf{Engagement of the meeting.}

\autoref{fig:engagement} shows the reported engagement points.
The average degree of engagement increases from 2.33 to 4.58 after the O-NEO is involved. 
The average degree of engagement of the P-NEO phase is 0.67 higher than that of the O-NEO phase.
A paired t-test shows a significant difference between the control phase and the O-NEO phase ($t_{11}=-4.700, p<0.01$).
Significance is also found in the difference between the control phase and the P-NEO phase ($t_{11}=-6.027, p<0.001$).
The result infers that NEO increases user engagement and helps the user to keep concentrating.

\textbf{Work load.}

The workload over the phase is shown in \autoref{fig:NASA_TLX}(a).
As most pairs of data in NASA-TLX violated the normality assumption, we utilized the Wilcoxon signed rank tests for significance analysis.
The mental demand decreases in the O-NEO phase ($Z=-2.388, p=0.017$) and the P-NEO phase ($Z=-2.701, p=0.007$), compared to the control phase.
The physical demand (PD) and the time demand (TD) decrease in the O-NEO phase (PD: $Z=-2.921, p=0.003$, TD: $Z=-2.462, p=0.014$) and the P-NEO phase (PD: $Z=-2.897, p=0.004$, TD: $Z=-2.825, p=0.005$) as well.
The performance rating of the P-NEO phase is significantly higher than that of the control phase ($Z=-2.366, p=0.018$).
These two phases also show lower requirements of effort (O-NEO: $Z=-2.101, p=0.036$, P-NEP: $Z=-2.589, p=0.010$) and lower frustration levels (O-NEO: $Z=-2.573, p=0.010$, P-NEO: $Z=-2.777, P=0.005$) compared with the control phase.
The difference between the O-NEO phase and the P-NEO phase fails to pass the significance analysis.

As for the workload of handling the disruptions (shown in \autoref{fig:NASA_TLX}(b)), similar differences are observed by the Wilcoxon signed rank tests between the two pairs: (1) the control phase vs. the O-NEO phase, (2) the control phase vs. the P-NEO phase.
Additional differences occur when comparing the effort and frustration levels between the O-NEO phase and the P-NEO phase.
The effort required in the P-NEO phase is lower than that of the O-NEO phase ($Z=-2.271, p=0.023$).
The frustration level is also lower in the P-NEO phase ($Z=-2.333, p=0.020$).
The mean NASA-TLX score of the O-NEO phase and the P-NEO phase is reported to be 2.90 and 2.56 on an equal-weighted 7-point scale, indicating that the general workload of handling disruptions with the help of NEO is relatively light.
These results imply that NEO has successfully cooperated with participants while they are occupied by something.
The P-NEO with personalized knowledge can better release participants from handling the disruptions than the O-NEO, which shows the value of personalization.

\begin{figure}[tb]
 \centering 
 \includegraphics[width=0.6\columnwidth]{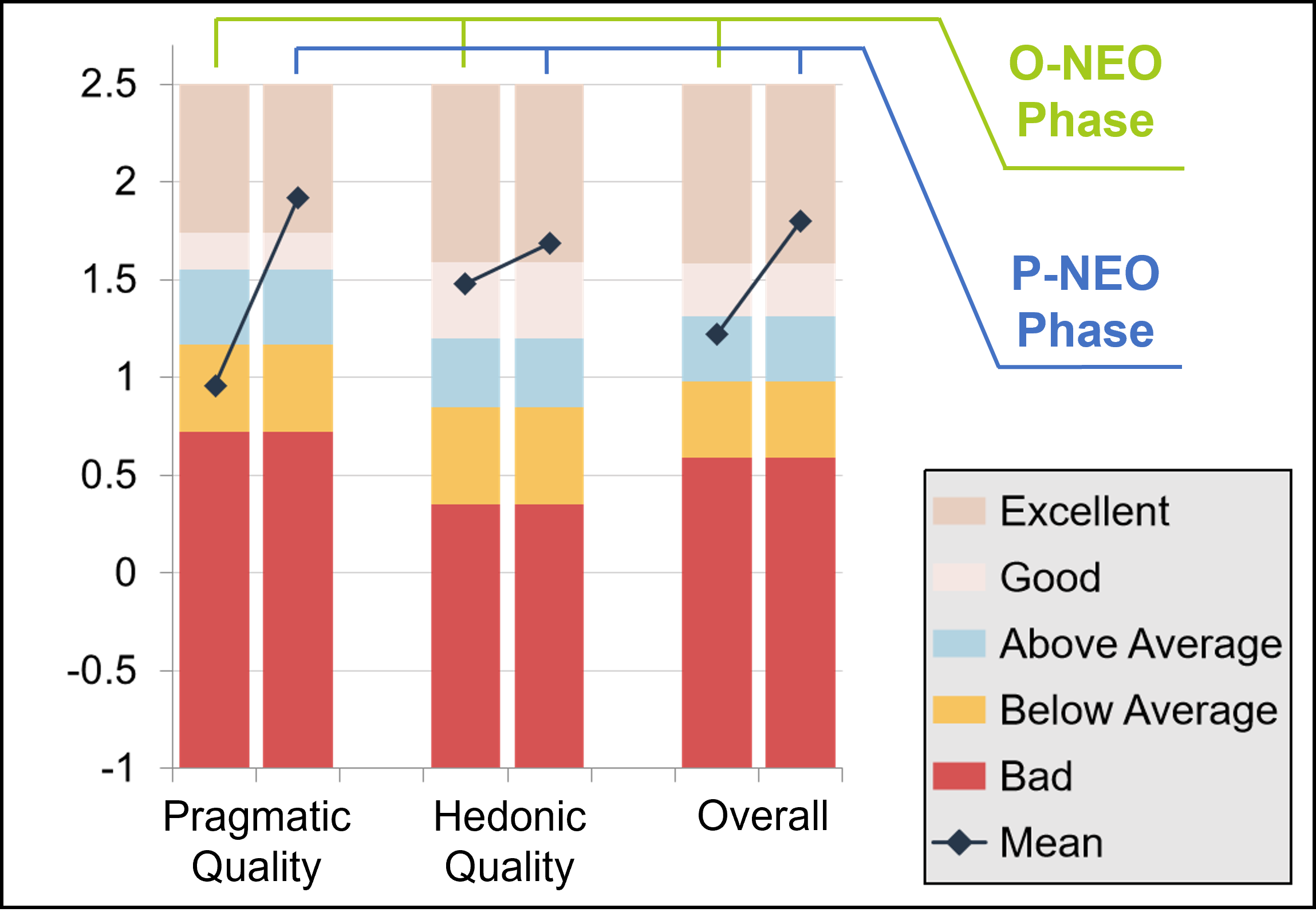}
 \caption{Results of the UEQ-S. In each group of bars, the left one indicates the result of O-NEO while the right one indicates the result of P-NEO.}
 \label{fig:UEQ}
\end{figure}

\begin{figure}[tb]
 \centering 
 \includegraphics[width=0.6\columnwidth]{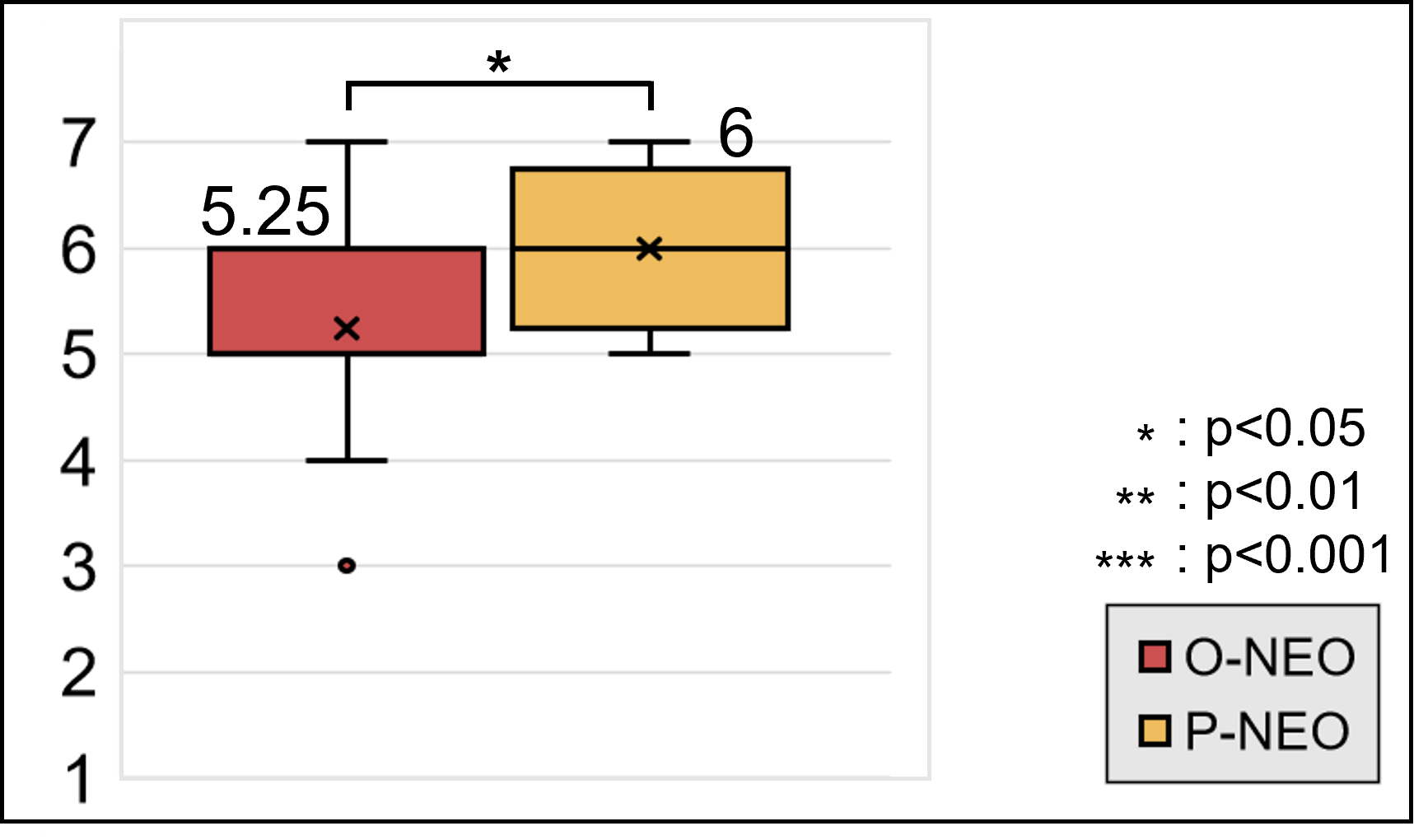}
 \caption{Satisfaction ratings of NEO.}
 \label{fig:satisfy}
\end{figure}

\textbf{User experience.}

The results of UEQ-S are shown in \autoref{fig:UEQ}.
The overall ratings are both above average, indicating that NEO has provided participants with a good experience of cooperation.
A significant improvement is captured in the channel of pragmatic quality (paired t-test, $t_{11}=-2.615, p=0.024$) after personalization.
We noticed that the average hedonic quality went up a level after personalization: from good to excellent.
The average of P-NEO's overall ratings (excellent) is two levels higher than that of O-NEO (good).

The average ratings of satisfaction that participants have reported for O-NEO is 5.25 (SD=1.06), while that for P-NEO reaches 6.00 (SD=0.74).
\autoref{fig:satisfy} visualizes the results.
The ratings for NEO significantly increase after personalization (Wilcoxon signed rank test, $Z=-2.041, p=0.041$).

\textbf{Semi-structured interview.}

Participants were asked to rate their will of using NEO in their daily lives on a 7-point scale (1=not at all, 7=very much so).
The average rating is 6.25 (SD=0.75), showing that participants hold a positive attitude towards NEO.
Participants' opinions about the design of NEO are summarized below.

(1) \textit{The way NEO observes and acts.}
The majority of the participants were satisfied with NEO.
``It can really reduce the disturbance,'' said P07.
``The design is simple but clever,'' agreed by P11, ``and it is pretty enough for me.''
Four expressed particular interest in the personalization function (P01, P06, P08-09).
P09 and P10 mentioned that answering the door was an important requirement and had been well handled by NEO.
P08 expressed the expectation of NEO to evaluate his attention and to offer water or food based on this evaluation: ``I don't want water if I'm in a debate, but I might want some when chatting.''
P09 emphasized the importance of personalization. ``I might not use NEO without personalization.'' said P09, ``I don't want a model designed by strangers.''

(2) \textit{Other embodiment for NEO. }
Four participants reported that the proposed embodiments were enough.
P01, P09, and P12 would like to have additional robotic arms to carry large objects or pick up books from a book shell.
Humanoid robots were nominated by two participants, while the idea of using drones was proposed by P06.
Different from other participants, P11, a student who majored in robotics, expressed his concerns: ``I think current embodiments are enough. Using other embodiment might cause additional danger.''

(3) \textit{More kinds of behavior. }
P03 and P06 both suggested extending the audio module of NEO so that NEO could not only receive the visitors but also chat with them.
The requirement of displaying caller information was addressed by P07, P09, and P12.
P02, P03, P06, and P07 all pointed out the need for assistance in taking notes or searching for information during the meeting.

(4) \textit{Ethical concerns. }
P05 expressed worries about the delay in personalization.
``I am afraid that NEO might open the door for my ex after we break up.'' said P05, ``When I interact with NEO, it seems like I'm looking at my past self.''
P06 reminded us that there might be issues if NEO observes confidential conversations.

\section{Discussion and future work} 

\subsection{System development}
\textbf{The meanings of joint observation. }
It is seen from \autoref{fig:sysexp1} that for most cases, the accuracies increase when more classes of sensors are activated.
We thus stress the necessity of joint observation for an agent to cooperate with people in both physical and virtual environments.
However, we also observe the difference in accuracy across cases.
There are cases that reported good accuracy results with only PU sensors, but poor accuracy with only PE or VE sensors.
This is caused by the difference in the number of cases that different sensor classes are related to.
For example, the PU class contains a rich amount of information that related to the user's occupation state and disruption states.
The VE class's information, on the other side, is related to fewer states than that of PU.
Although involving more cases in the test set could reduce such differences, we argue that the listed cases already reflect the advantage of joint observation.
There are also some cases that are not parsed by NEO, such as ``knocking over a glass of water''.
This is caused by the limitation of NEO's current sensor set and could be further explored by extending sensor types for NEO.

\textbf{Compatibility}
The results of Stage-2 demonstrate NEO's compatibility of cooperating with different kinds of people.
Moreover, the results of Stage-1 show the compatibility of NEO to work with different sensor settings.
This means that users can activate a subset of NEO's sensor set based on their needs and have NEO working without additional development.
It shows NEO's potential of being implemented in conventional rooms.
The users do not have to implement the whole system at the beginning.

\subsection{User experience}
\textbf{Human-agent symbiosis.}
In the user study, we put our participants in a hectic setting where they must handle several things at once.
Then, we contrasted their workload and level of involvement with and without NEO's assistance.
It is concluded from the results that NEO did notice the disruption, made the right decision, and reduced the distraction through bi-environmental action.
The feedback from participants through the UEQ-S and the semi-structured interview also suggests that NEO is helpful during the VR meeting, especially after personalization.
Regarding the two physiological disruptions, although NEO has brought water and food to the user upon detection of thirst and hunger, saving their effort to take off the HMD, drinking and eating are still distracting to the immersive experience.
We conclude from the above evidence that NEO takes one more step towards the two expectations of a man-computer symbiosis~\cite{licklider1960man} to provide users with a not only viable but also productive partnership.
The interview result also implies that NEO is well accepted among the participants.
These conclusions picture a bright future of implementing NEO into homes as an AI partner.

\textbf{Personalization.}
From another perspective, the results of UEQ-S show that the personalized NEO is preferable to the original one.
The difference between the UEQ-S result of O-NEO and P-NEO supports the necessity of personalization.
P09 also stressed its importance in the interview.
Combining these with reports from the interview that the P-NEO is more popular among the participants, we conclude that NEO did learn something from the participants during the simple feedback interaction session.
The overall user experience of P-NEO is excellent, demonstrating the effect of personalization.
However, as mentioned by P05, there is a delay between the feedback session and the actual change of behaviour.
This delay may cause issues when sudden changes in user preferences happen.

\subsection{Application}
Systems like NEO could be implemented in more smart home scenarios. For example, it could be used in kitchens to give warnings when the food is about to burn; it could also detect pets if they enter the working area and warn the user, with extensions in the observation module.
NEO could also be used in other scenarios such as health care. 
Disabled patients may immerse in virtual worlds for entertainment in the future. 
In this case, NEO could help maintain their physiological needs (bring food or medicines) or ensure their medical devices are running properly (notify medical staff when an infusion completes).
Additionally, in manufacturing scenarios, if an accident occurs in a factory but unnoticed by humans, NEO could detect it through joint observation and handle the fault point.

\subsection{Future work}
With the capability and usability of NEO proved by these experiments, we are confident to integrate more sensors and algorithms into the system.
Following the suggestions given by the participants, we would like to update NEO's set of embodiments with robotic arms and grippers in the future.
The method of providing visitors information could be further improved to better align with the ongoing virtual scene.

We address the case of one physical environment with one virtual environment in this work. One physical environment corresponding to multiple virtual environments is a special case in SR. If there is only one user, NEO could handle these cases when the user switches from one virtual environment to another. If there are multiple users immersed in multiple virtual environments, NEO should be linked to each environment, and the collected data will be processed in the backend and integrated by the user. Such questions should be addressed in future work.

\section{Conclusion}
In this paper,  we discuss methods for designing an agent under the topic of human-computer symbiosis.
We propose a method that consists of three points: joint observation, scalable decision, and joint action.
To verify the feasibility of this method, we develop a prototype system named NEO, which cooperates with people while they are immersed in a virtual environment.
Two system experiments and a user study are conducted to evaluate this system.
The results support the idea of joint observation and joint action.
They also demonstrate the usability and effectiveness of NEO.
Overall, NEO represents a promising step forward in the development of agents that can effectively operate across physical and virtual environments, facilitating human-agent interaction and providing users with personalized and responsive support, with important implications for future research in this field.

\bibliographystyle{abbrv}



\end{document}